\documentclass[aps,letterpaper,twocolumn,nofootinbib]{revtex4}

\pdfoutput=1

\usepackage{amsthm}         
\usepackage{graphicx} 	    
\usepackage{amsmath} 	   
\usepackage{amssymb}       
\usepackage{amsfonts}
\usepackage{setspace}	    
\usepackage{capt-of}

\begin{document}

\title{Simulation of Black Hole Collisions in Asymptotically AdS Spacetimes}

\author{Hans Bantilan}
\author{Paul Romatschke}
\address{Department of Physics, University of Colorado Boulder, CO 80309}

\begin{abstract}
We present results from the evolution of spacetimes that describe the merger of asymptotically global AdS black holes in 5D with an SO(3) symmetry. Prompt scalar field collapse provides us with a mechanism for producing distinct trapped regions on the initial slice, associated with black holes initially at rest. We evolve these black holes towards a merger, and follow the subsequent ring-down. The boundary stress tensor of the dual CFT is conformally related to a stress tensor in Minkowski space which inherits an axial symmetry from the bulk SO(3). We compare this boundary stress tensor to its hydrodynamic counterpart with viscous corrections of up to second order, and compare the conformally related stress tensor to ideal hydrodynamic simulations in Minkowski space, initialized at various time slices of the boundary data. Our findings reveal far-from-hydrodynamic behavior at early times, with a transition to ideal hydrodynamics at late times.  
\end{abstract}

\maketitle

\noindent{\bf{\em Introduction ---}}
At the Relativistic Heavy Ion Collider (RHIC) and the Large Hadron Collider (LHC), the collision of heavy ions 
at relativistic speeds results in the formation of strongly-coupled matter consisting of deconfined quarks 
and gluons. Although the system expands and cools quite rapidly, experimental signatures indicate that 
strong interactions between partons are sufficient to locally thermalize the system within $10^{-23}$ seconds 
(1 fm/c), creating a state of quantum matter known as the Quark Gluon Plasma (QGP). The plasma is well described 
by ideal hydrodynamics for a time of several fm/c before it reaches the QCD phase transition. In view of 
the AdS/CFT correspondence~\cite{Maldacena:1997re,Gubser:1998bc,Witten:1998qj}, it has been 
proposed~\cite{Nastase:2005rp,Amsel:2007cw,Grumiller:2008va,Gubser:2008pc,Albacete:2008vs,Aref'eva:2009wz} 
that this picture of QGP formation may lend itself to a description in terms of black holes (BH) in 
asymptotically anti-de Sitter (AdS) spacetimes. For a recent and comprehensive review, 
see~\cite{DeWolfe:2013cua}. Head-on BH collisions are well understood to result 
in a single remnant BH following the formation of an encapsulating apparent horizon. At sufficiently high boosts, 
this result remains true for the head-on collision of other compact objects~\cite{Choptuik:2009ww,East:2012mb,Rezzolla:2012nr}. In the asymptotically AdS non-rotating case, the resulting 
spacetime settles down to the AdS-Schwarzschild solution, whose horizon area is identified with the temperature 
of the boundary conformal field theory (CFT). The way this remnant BH settles down could provide important details 
of how a non-equilibrium state in the boundary quantum theory evolves and eventually thermalizes. A merger of BHs 
in the bulk provides a dynamical way of arranging for these non-equilibrium states.

In studies to date, symmetries have been imposed on the bulk spacetime solution to make the 
gravitational problem in AdS more tractable, which in turn imply particular symmetries on the boundary flow. 
In~\cite{Chesler:2010bi}, colliding shockwaves in AdS$_5$ were studied in $2+1$ dimensional gravity 
simulations using a scheme based on characteristic evolution, leading to $1+1$ boundary dynamics. 
In~\cite{Heller:2011ju}, the study focused on $1+1$ dimensional bulk solutions, making use of an evolution 
scheme based on the ADM decomposition of the Einstein field equations. The boundary flows obtained from 
these gravity solutions was compared to the flow one expects from hydrodynamics. This comparison was taken 
further in~\cite{vanderSchee:2013pia}, where gravity data was used as initial data for relativistic hydrodynamics. 
Recently, significant effort has been put into constructing Minkowski flows with dynamics in more than one 
boundary spatial coordinate. For instance, in~\cite{Fernandez:2014fua} a Minkowski flow with a fixed but 
non-trivial transverse profile was studied using linear gravity corrections in a $2+1$ dimensional bulk 
spacetime background.

An important long-term goal of this program is a holographic description of general heavy-ion collisions 
with an arbitrary impact parameter, which requires fully 4+1 dimensional black hole mergers in the asymptotically 
AdS$_5$ setting. The main purpose of this paper is to present the first proof-of-principle simulation of 
stable black hole mergers in AdS$_5$. We preserve an SO(3) symmetry in the bulk so that the flows we 
construct in Minkowski space, which can be described by Cartesian coordinates $(t^\prime,x_1,x_2,x_3)$, 
exhibit dynamics that depend on $x_3$, $\sqrt{x_1^2 + x_2^2}$, and $t^\prime$ despite having started from 
an effective $2+1$ bulk evolution. The origin of the enhanced spatial dependence in our Minkowski flows stems from 
the conformal relation between Minkowski space and the $\mathbb{R} \times S^3$ boundary of the 
global AdS$_5$ bulk i.e. the flows we obtain in Minkowski space are generated by gravitational dynamics in 
the bulk, along with a piece that is of a purely conformal origin. This approach was first considered 
in~\cite{Friess:2006kw} for linearized perturbations around the AdS-Schwarzschild solution, and further 
clarified in~\cite{Gubser:2010ze,Gubser:2010ui}. Simulations of the ring-down of highly distorted BHs in the 
fully non-linear regime were obtained in~\cite{Bantilan:2012vu}. In that study, the gravity solutions 
were found to correspond to boundary dynamics that is completely described by second order viscous hydrodynamics. 
What is novel in the current work is a move away from the regime were hydrodynamics dominates 
from the outset, by starting with a gravitational slice that contains distinct trapped regions that move towards 
a merger. In this way, the post-collision state of the spacetime arises dynamically, rather than being put 
in by hand. In the following, we work in geometric units where Newton's constant $G$ and the speed of light 
$c$ are set to $1$. We also set the AdS length scale $L$ to $1$, particularly when quoting explicit 
numerical results for the boundary CFT stress tensor.

\noindent{\bf{\em Numerical Scheme ---}}
The results described here were obtained using a numerical code first presented in~\cite{Bantilan:2012vu} that 
simulates asymptotically global AdS spacetimes by solving the Einstein field equations 
with a negative cosmological constant $\Lambda_5=-6/L^2$, using the generalized harmonic (GH) evolution 
scheme introduced in~\cite{Pretorius:2004jg,Pretorius:2005gq}. 
This formalism is based on spacetime coordinates $x^\mu$ that each satisfies a scalar wave equation 
$\square x^{\mu}=H^{\mu}$. In the context of asymptotically AdS spacetimes, demanding that the desired 
metric asymptotics be preserved during the bulk evolution necessitates specific relations amongst the 
leading-order behavior of the metric $g_{\mu\nu}$ and the GH source functions $H_\mu$. In terms of the 
global boundary coordinates 
$x^m=(t,\chi,\theta,\phi)$ and the AdS radial coordinate $\rho$, if we write the metric as
$g_{mn}      =g^{AdS}_{mn}       + (1-\rho^2)\bar{g}_{mn}$,  
$g_{\rho\rho}=g^{AdS}_{\rho\rho} + (1-\rho^2)\bar{g}_{\rho\rho}$,
$g_{\rho m}  =g^{AdS}_{\rho m}   + (1-\rho^2)^2\bar{g}_{\rho m}$, 
and the source functions as
$H_m    = H^{AdS}_m + (1-\rho^2)^3 \bar{H}_m$,
$H_\rho = H^{AdS}_\rho + (1-\rho^2)^2 \bar{H}_{\rho}$, 
then the following set of algebraic relations enforced at the $\rho=1$ boundary are sufficient to yield 
stable evolution: 
$\bar{H}_m |_{\rho=1}    = \frac{5}{2} \left. \bar{g}_{\rho m} \right|_{\rho=1}$,
$\bar{H}_\rho |_{\rho=1} = 2 \left. \bar{g}_{\rho \rho} \right|_{\rho=1}$.
In the context of black hole mergers in AdS, we choose a gauge that spatially 
interpolates from this boundary gauge to a gauge in the bulk that is harmonic with respect to pure 
AdS i.e. $\bar{H}_\mu = 0$. We found that extending the boundary gauge into the bulk in this way helps 
to control coordinate singularities that tend to develop in the post-merger evolution near the apparent horizon. 
These are particularly evident in the lapse function $\alpha=-( 1/\sqrt{-g^{tt}} )$, 
and we found it helpful in some cases to include lapse-damping terms. These terms appear in the 
GH source functions as $\kappa\left( n_\mu -\bar{n}_\mu \right)$, where $\kappa$ is a constant, 
$n_\mu dx^\mu = -\alpha dt$ is the unit normal to slices of constant $t$, and $\bar{n}_\mu dx^\mu$ encodes 
the choice of lapse-damping. We make the choice 
$\bar{n}^\mu = \left( \partial_t \right)^\mu/\alpha + n^\mu\log\alpha$, 
first considered in~\cite{Lindblom:2009tu}. 

We use the excision method to excise a proper subset of each trapped region on the grid.  
This method removes the physical singularities associated with black holes from the computational 
domain, while taking advantage of the causal structure of spacetime within trapped regions so that 
no boundary conditions are placed at the excision surface. 
We search for apparent horizons using the flow method described 
in~\cite{Pretorius:2004jg}, keeping track of several trapped regions simultaneously. 
To generate trapped surfaces on the initial time slice, we make use of a free massless scalar field 
$\varphi$ with a spatial profile on the initial slice in the form of a Gaussian 
$\varphi(\rho,\chi) = (1-\rho^2)^4 A_0 \exp\left( -R(\rho,\chi)^2 / \delta^2 \right)$
where
$R(\rho,\chi) = \sqrt{\left(x(\rho,\chi)^2-x_0\right)/{w_x}^2 + \left(y(\rho,\chi)^2-y_0\right)/{w_y}^2}$, 
$x(\rho,\chi) = \rho \cos\chi$, and $y(\rho,\chi) = \rho \sin\chi$. We choose $A_0$ sufficiently 
large and $\delta$ sufficiently small to ensure that the initial time slice contains trapped regions by virtue 
of prompt scalar field collapse. We arrange for two black holes to start from rest at antipodal points 
on the axis, evolving towards a collision at the origin, by choosing the $\varphi$ spatial 
profile on the initial slice to comprise of two Gaussians centered at $x_0=\pm D$, $y_0=0$.

\begin{figure*}[t]
	\includegraphics[width=0.48\linewidth]{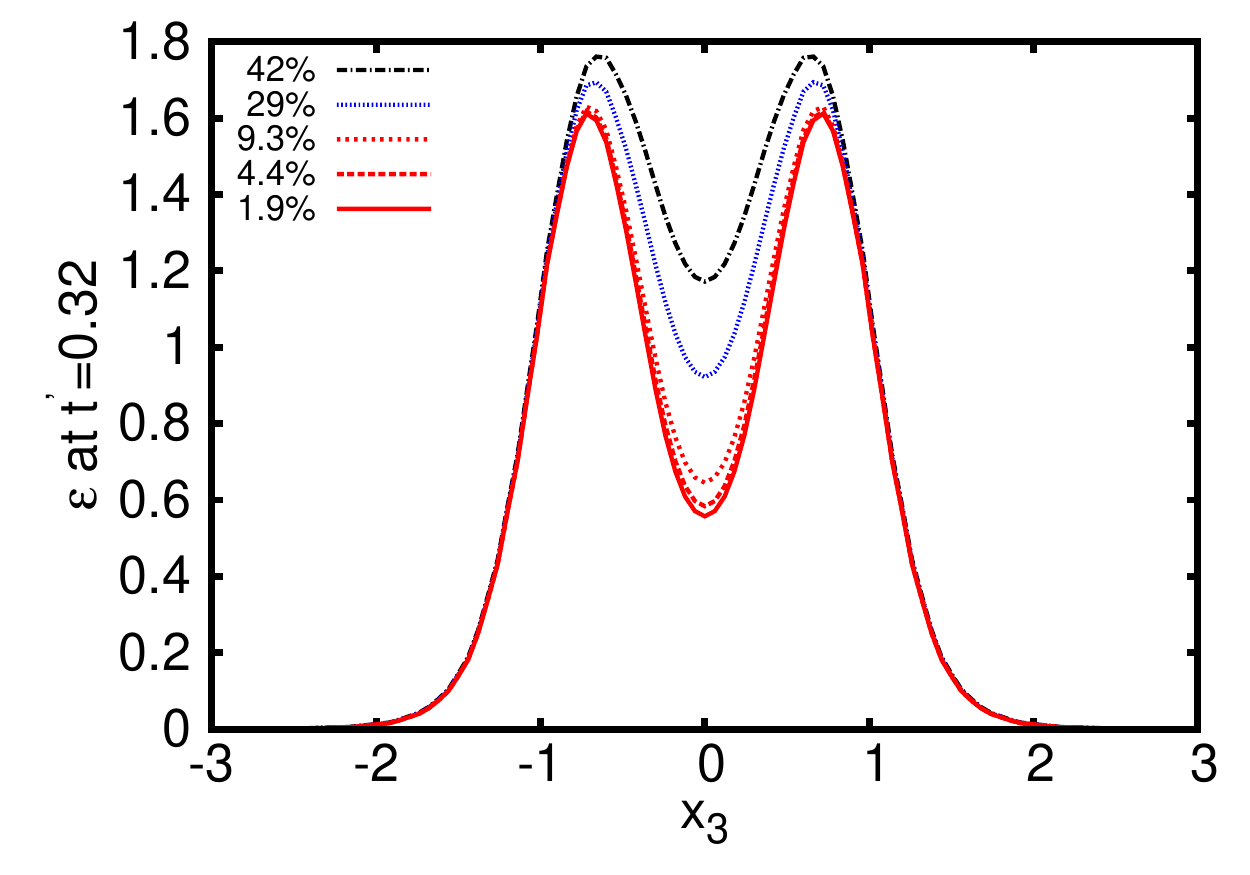}
    \hfill
	\includegraphics[width=0.48\linewidth]{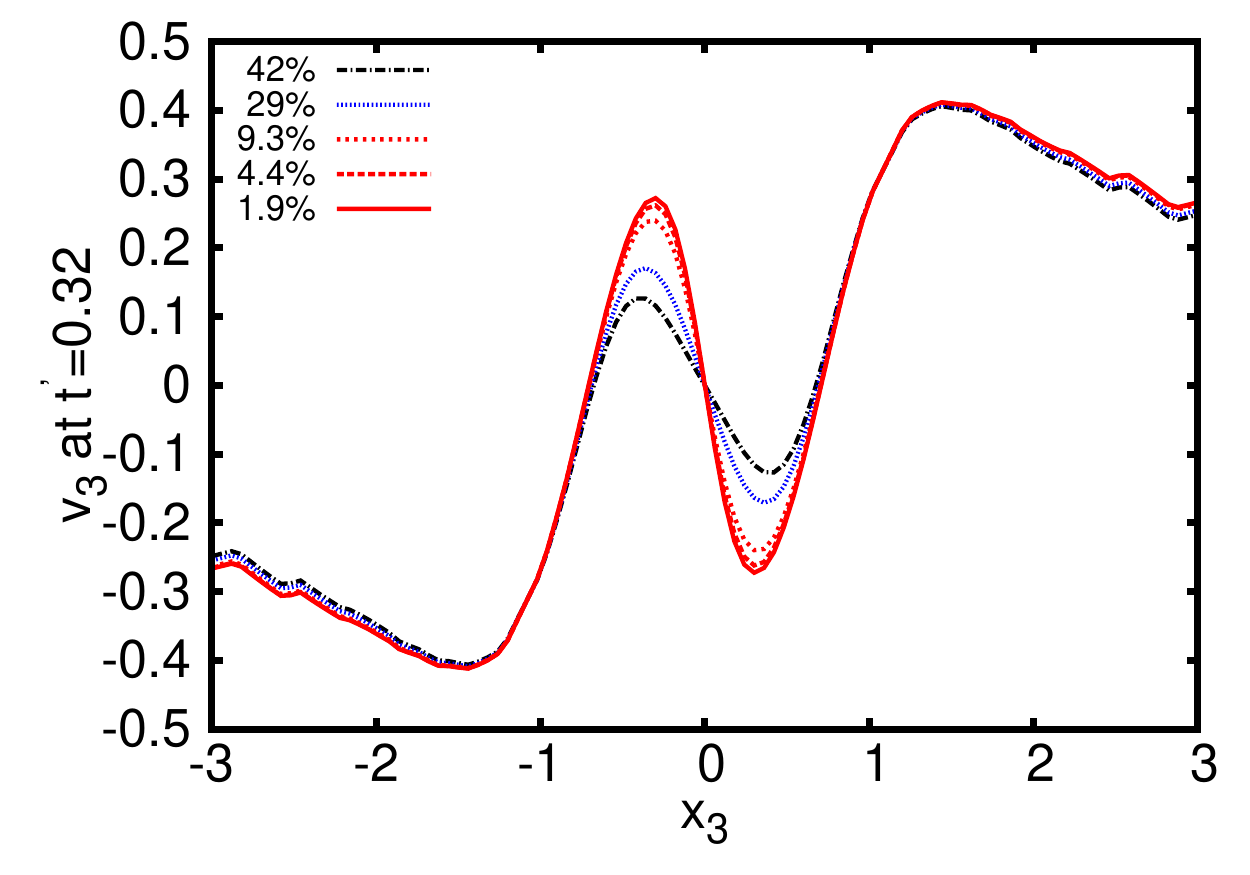}
{\caption{The effect of changing the size of the regulator BH as seen from the $x_1=x_2=0$ spatial slice of 
Minkowski space at $t^\prime=0.32$. Here, the mass of the regulator BH is expressed as percentage of the total 
spacetime mass. Left: The energy density distribution along $x_3$. Right: The $x_3$-component of three-velocity 
along $x_3$.}\label{fig:multi}}
\end{figure*}

\begin{figure*}[t]
  \includegraphics[width=0.48\linewidth]{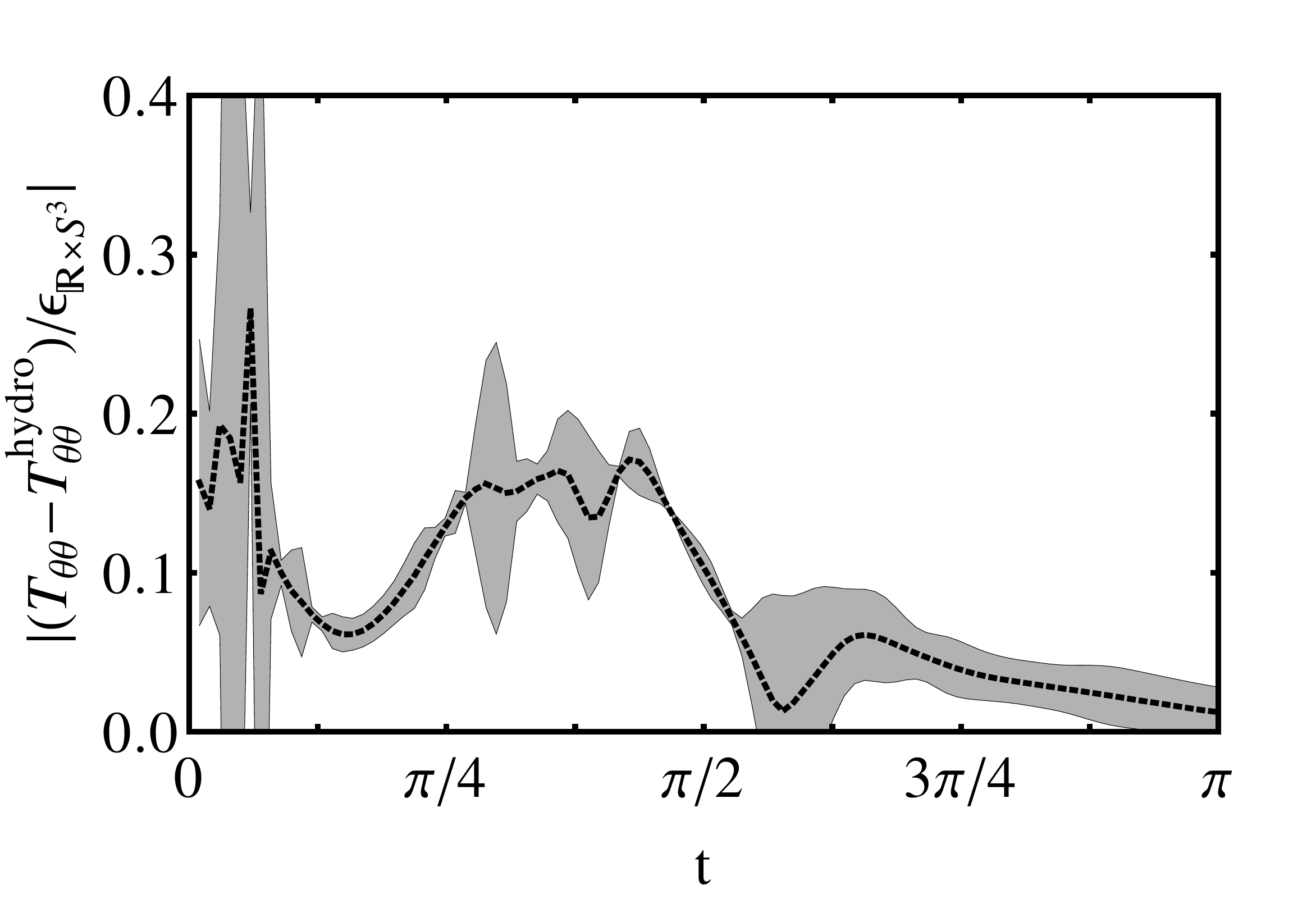}
  \hfill
  \includegraphics[width=0.48\linewidth]{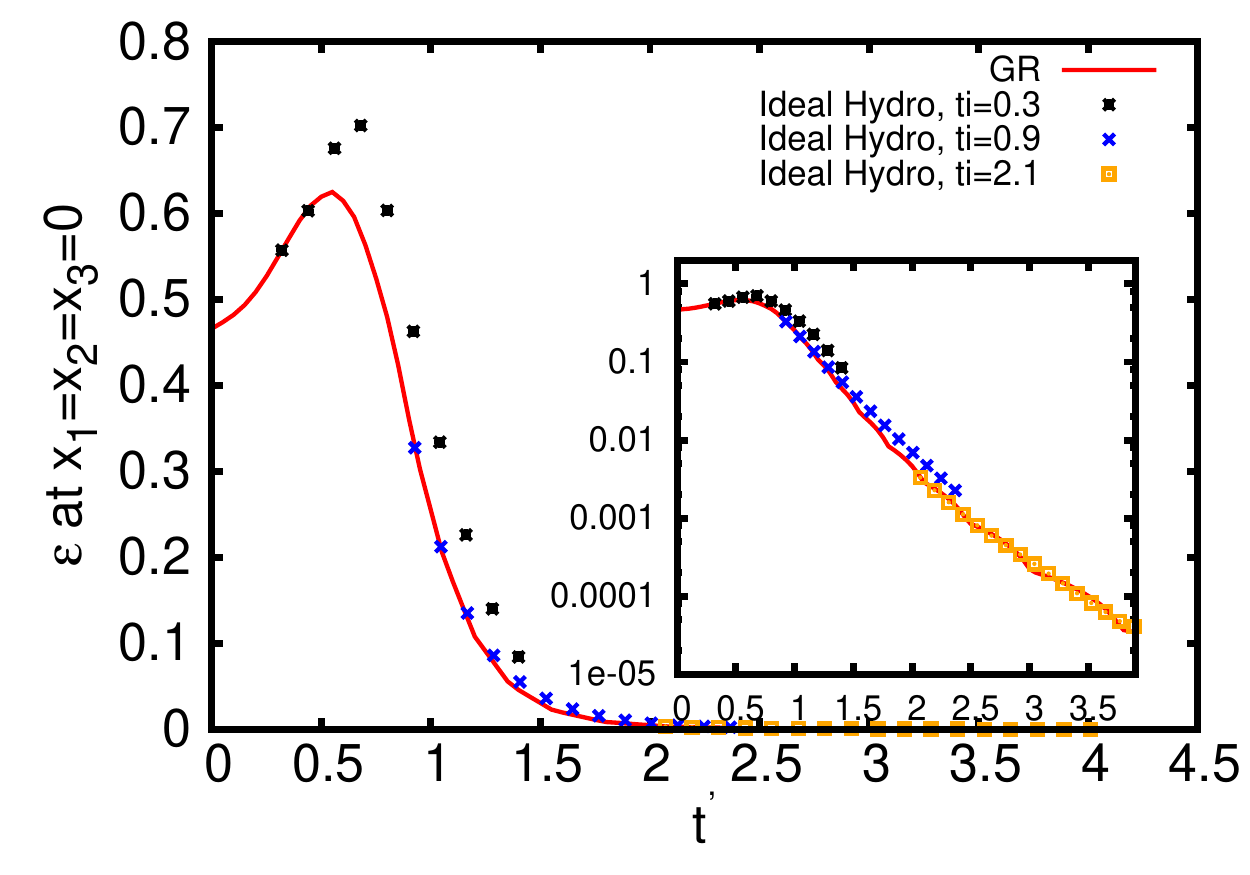}
  \caption{Left: Comparison on $\mathbb{R} \times S^3$, showing the normalized mismatch between the boundary 
stress tensor and its hydrodynamic counterpart, with viscous corrections of up to second order. Here, $|.|$ 
denotes an $L^2$ norm over $\chi \in [0,\pi]$; for more details, see~\cite{Bantilan:2012vu}, where a similar 
analysis was performed on distorted AdS$_5$ black holes. Data from the highest resolution run (dashed black line) 
is displayed with an estimated uncertainty (grey shaded region). Right: Comparison on $\mathbb{R}^{3,1}$, 
showing the time evolution of energy density at the origin, with data from the boundary stress tensor of 
the gravity solution (``GR'') and from ideal 3+1 dimensional relativistic hydrodynamics (``Ideal Hydro''). 
For computational reasons, a small but non-vanishing viscosity was used in the hydrodynamic simulation. 
The hydrodynamics is initialized with gravity data at $t^\prime_i=0.3, 0.9, 2.1$.
}\label{fig:rxs3_minkowski}
\end{figure*}

\begin{figure*}[t!]
\centering
\includegraphics[width=1.0\linewidth,angle=0]{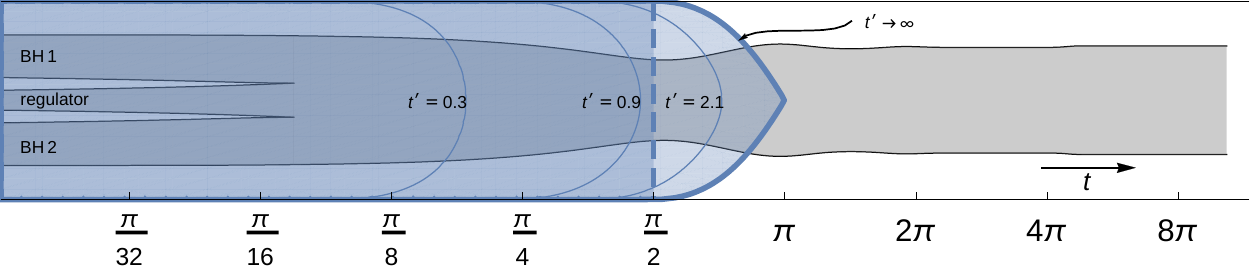}
\caption{Spacetime diagram of the bulk solution for a BH-BH simulation with a $1.9\%$ mass regulator BH. 
The vertical is parametrized by $\rho\cos\chi$ for $\rho \in \left[0,1\right]$ and $\chi \in \left[0,\pi\right]$, 
and the horizontal is parametrized by bulk global time $t$. The apparent horizon (solid black line) bounds each 
trapped region (grey shaded region). At early times $t<\pi/16$, there are three distinct trapped regions 
corresponding to two physical BHs and a regulator BH centered at the origin. The physical BHs move towards 
each other from rest and eventually merge with the regulator BH around $t\approx\pi/16$. The resulting 
post-merger BH rings down and eventually becomes quiescent at late times $t\gtrsim\pi$. The Poincar\'e patch 
(blue shaded region) is the wedge shaped region that tapers up towards $t=\pi$. The boundary of the 
Poincar\'e patch is where we extract all CFT quantities.
}\label{fig:bhdiagram}
\end{figure*}

\begin{figure*}[t!]
        \centering
        \includegraphics[width=0.8\linewidth,angle=0]{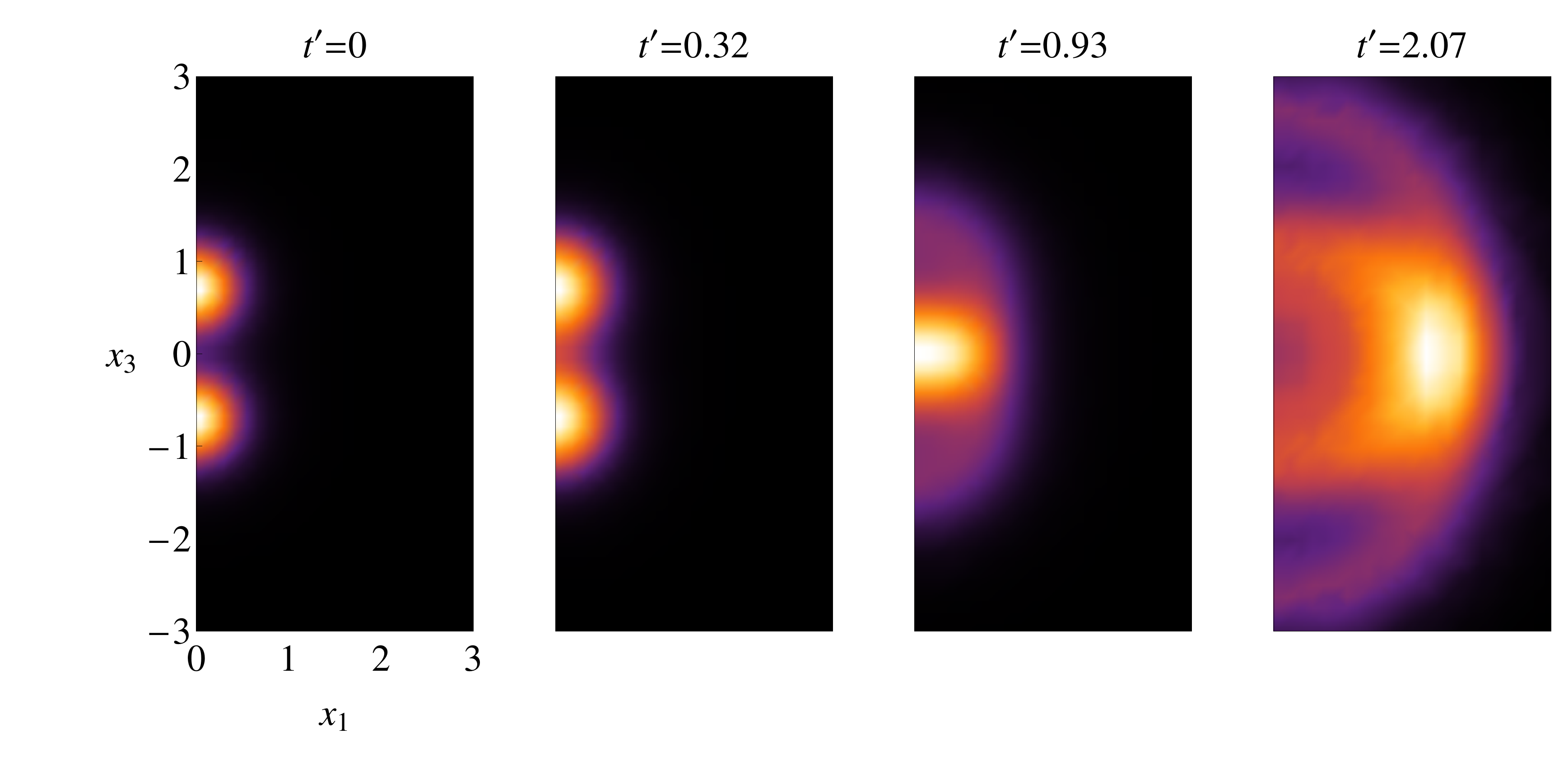}
\caption{ Snapshots of the energy density $\epsilon_{\mathbb{R}^{3,1}}$ in Minkowski space for a BH-BH 
simulation with a $1.9\%$ mass regulator BH, linearly rescaled such that all panels attain the same 
numerical maximum, to emphasize qualitative features of the flow. The four different panels contain the 
spatial profile of energy density in the $x_1 - x_3$ plane at four different constant $t^\prime$ slices. 
One can recover the full spatial dependence of the flow by simply rotating each of these panels about the 
$x_3$ axis.
}\label{fig:minkowskiplot}
\end{figure*}

\noindent{\bf{\em Hydrodynamic Comparison ---}}
In this section, we describe two different methods for quantifying the hydrodynamic behavior of the 
boundary stress tensor obtained from our bulk gravity solutions. The first method is an analysis on 
$\mathbb{R} \times S^3$, which involves verifying that the conformal constitutive relations hold, 
extracting the energy density and four-velocity from the boundary stress tensor, then, 
using these ingredients, reconstructing the boundary stress tensor's hydrodynamic counterpart order by order 
in the velocity gradient expansion. Here, we implement this method using viscous correction terms of up to second 
order, calculated in~\cite{Loganayagam:2008is,Baier:2007ix}. The second method we use is an analysis 
on $\mathbb{R}^{3,1}$, which involves mapping the boundary stress tensor of the global AdS bulk to 
a conformally related stress tensor in Minkowski space. The time evolution of this stress tensor is then 
compared to an ideal 3+1 hydrodynamic simulation performed using the relativistic Lattice Boltzmann solver 
presented in~\cite{Romatschke:2011hm}, initialized with data from a constant Minkowski time slice of the 
gravity solution. 

In Minkowski space, we restrict our analysis to a comparison with ideal hydrodynamics, 
with the intention of investigating the effect of viscous corrections in a future study. 
To see why this is a reasonable baseline comparison, particularly at late times, 
it is instructive to consider the solution in the bulk that all our solutions eventually approach: 
a static AdS-Schwarzschild BH, the boundary of which describes a flow on $\mathbb{R} \times S^3$ 
with constant energy density $\epsilon_{\mathbb{R} \times S^3}$ and four-velocity 
$u_{\mathbb{R} \times S^3}^\mu = (1,0,0,0)$. On $\mathbb{R}^{3,1}$, this flow is mapped onto a time varying 
and spatially dependent energy density $\epsilon_{\mathbb{R}^{3,1}}=\epsilon_{\mathbb{R}\times S^3}/W^4$ 
and four-velocity $u_{\mathbb{R}^{3,1}}^a = (u^0,x_1 t^\prime/W,x_2 t^\prime/W,x_3 t^\prime/W)$, 
where $W=\sqrt{(t^\prime)^2+(1+x_1^2+x_2^2+x_3^2-(t^\prime)^2)^2/4}$ and 
$u^0=\sqrt{1+(x_1 t^\prime/W)^2+(x_2 t^\prime/W)^2+(x_3 t^\prime/W)^2}$. 
This example is particularly illustrative because it cleanly separates the contribution of the bulk gravity 
dynamics from the dynamics generated by the conformal mapping: for this solution, the dynamics on 
$\mathbb{R}^{3,1}$ is entirely from the conformal mapping. Although this flow exhibits 
non-trivial velocity gradients, it is nevertheless {\em exactly} inviscid. This follows from the fact that 
all viscous correction terms transform homogeneously under the conformal mapping from 
$\mathbb{R} \times S^3$, where all these terms vanish identically. Thus, at late times when our solutions 
have begun to approach the AdS-Schwarzschild BH solution, one expects the dynamics in $\mathbb{R}^{3,1}$ to 
be well described by solutions of the ideal hydrodynamic equations. 

\noindent{\bf{\em BH-BH Collisions ---}}
We now perform the analysis described in the previous section on the full BH-BH bulk solutions. 
We choose initial conditions with scalar field parameters $D=0.5$, $A_0=1.25$, $\delta=0.025$, $w_x=w_y=1$. 
The centers of the two BHs that form via prompt scalar field collapse are initially separated by a distance 
$2 L$ in the bulk. Because of our choice of polar-like coordinates, we found it necessary 
to additionally introduce a regulator BH centered at the bulk origin, in order to side step the 
severe restriction on step size imposed by the Courant-Friedrichs-Lewy (CFL) condition near the origin: our 
simulations thus describe triplet BH systems. We study the effect of this regulator BH on the 
boundary stress tensor, and find that we can engineer it to be small enough that its effect on the 
stress tensor evolution becomes negligible, particularly when its mass is made to be 
less than $10$ percent of the total spacetime mass; cf. Fig.~\ref{fig:multi}. The results of the comparison 
to hydrodynamics on $\mathbb{R} \times S^3$ for a representative component of the boundary stress tensor is 
shown in the first panel of Fig.~\ref{fig:rxs3_minkowski}. At early global times $t$, the solution converges 
to behavior that is different from second order viscous hydrodynamics by up to $\approx 20\%$.  At late $t$, 
the mismatch between the solution and second order viscous hydrodynamics is essentially zero 
to within truncation error. 

From an $\mathbb{R}^{3,1}$ perspective, this BH-BH system consists of two energy density lumps on the 
$x_3$-axis, each of which disperses in all directions. The resulting wave fronts have 
a maximum overlap at the origin at about $t^\prime \approx 0.5$, which then continue to disperse with 
energy density falling off roughly as $\sim 1/(t^\prime)^4$ at late times. A spacetime 
diagram of the global AdS$_5$ bulk that summarizes this solution's dynamical horizon is shown in 
Fig.~\ref{fig:bhdiagram}, and various snapshots of the energy density in Minkowski space are shown in 
Fig~\ref{fig:minkowskiplot}. The time evolution of energy density $\epsilon_{\mathbb{R}^{3,1}}(t^\prime)$ at 
the origin is shown in the second panel of Fig.~\ref{fig:rxs3_minkowski}, along with a comparison to ideal 
hydrodynamic evolution initialized with gravity data at three different Minkowski times $t^\prime$. 
Ideal hydrodynamics initialized at an early time slice e.g. $t^\prime=0.3$ overshoots the maximum central 
energy density of the gravity data, and continues to evolve at late times to energy densities that 
are consistently larger than those extracted directly from gravity. When initialized at a sufficiently late 
time e.g. $t^\prime=2.1$, the fluid-dynamical evolution recovers good agreement with the energy density 
evolution of the gravity data, implying that the bulk solution has entered the ideal hydrodynamic regime. 

\noindent{\bf{\em Conclusion ---}}
We have presented the first successful numerical simulation of BH mergers in asymptotically 
AdS$_5$ spacetimes. We extracted the stress tensor of the dual CFT corresponding to these mergers in the bulk, 
finding non-equilibrium behavior at early times and close-to-equilibrium 
evolution given by ideal hydrodynamics at late times. At present, we are still limited to modest 
initial separations of the BH collision participants, primarily due to the difficulty in finding a gauge in 
the bulk that leads to stable post-merger evolution. We believe that our work constitutes an important step 
towards creating ``realistic'' models of the early, far-from equilibrium stages following the collision 
of heavy-ions. An imminent future effort is to relax the symmetry in the bulk, to obtain flows that are 
able to make closer contact with the real-world flows observed at RHIC and the LHC.

\noindent{\bf{\em Acknowledgments ---}}
We would like to thank Frans Pretorius, Steven Gubser, and Luis Lehner for many valuable discussions. 
We gratefully acknowledge support from the Department of Energy, 
DOE awards DE-SC0008027 and DE-SC0008132. Simulations were run on the {\bf Eridanus} cluster at 
the University of Colorado Boulder and the {\bf Orbital} cluster at Princeton University.

\bibliography{final_oct_29_2014}

\end{document}